\documentclass{article}

\usepackage{graphicx}
\usepackage[round]{natbib}
\usepackage{psfig}
\usepackage{epsfig}

\usepackage{graphicx,amsmath,amsfonts,latexsym}

\newcommand{\var}{{\rm Var}}

\newcommand{\mse}{{\rm MSE}}

\title{
Year ahead prediction of US landfalling hurricane numbers:
the optimal combination of long and short baselines
}

\author{Stephen Jewson\footnote{\emph{Correspondence email}: \texttt{x@stephenjewson.com}},
Christopher Casey and Jeremy Penzer\\}

\begin{document}
\maketitle

\begin{abstract}
Annual levels of US landfalling hurricane activity averaged over the last 11 years
(1995-2005) are higher than those averaged over the previous 95 years (1900-1994). How, then, should we
best predict hurricane activity rates for next year? Based on the
assumption that the higher rates will continue we use an optimal combination of averages
over the long and short time-periods
to produce a prediction that minimises MSE.
\end{abstract}

\section{Introduction}

There is considerable commercial interest in the prediction of
hurricane activity, although different industries are interested
in predictions over different time-scales. The insurance industry,
for instance, is mostly interested in the year-ahead timescale,
since year-ahead forecasts allow time for insurance rates to be adjusted
appropriately. Motivated by this the purpose of this article is to investigate some of
the properties of such year-ahead predictions.
A complete year-ahead prediction would consider all aspects of hurricane
activity such as hurricane intensity, size,
timing, location, and so on. In this study, however, we will focus
on one aspect only: the prediction of the annual number of hurricanes.

One might divide methods for the year-ahead prediction of hurricane numbers
into \emph{model-free} and \emph{model-based} methods.
Model-free methods do not make any assumptions about what has driven
variations in historical hurricane numbers. For example, the study described in~\citet{j81}
only makes the assumption that methods for hurricane
prediction that have worked well in the past will work well in the future.
The problem of finding good prediction methods then reduces to finding methods
that would have worked well in the past.

Model-based methods, on other hand, make much more specific assumptions,
of the form of a model for what has happened in the past and what will happen
in the future. The problem of finding good prediction methods then reduces
to (a) choosing a good model and (b) understanding how best to make
predictions, given the model.

It is not, in general, possible to say which of model-free or model-based
methods is better. The model-free method cited above is elegant because of the
minimal assumptions, but may not work well if the future is very unlike the past.
Model-based methods can always be criticized for the
particular choice of model, which is always arbitrary, and indeed always wrong (at some level),
but they give more flexibility in terms of incorporating non-stationarities, and
other assumptions. They are ideal for investigating how different assumptions
lead to different conclusions.

In this particular study, we investigate the possibility of using a model-based
method for predicting the number of landfalling hurricanes for next year.
We use one of the simplest possible non-trivial models. Our purpose is not actually
to make accurate hurricane predictions per se (since one would probably want to use
a more complex model for that) but to illustrate the interaction
between models and predictions based on those models, and, in particular,
to highlight the trade-off between bias and error variance in this problem.

\section{Unbiased and biased categorical forecasts}

When making a forecast, it makes sense to state precisely what that forecast
is trying to achieve, in terms of the error statistics of the forecast.
For single-value forecasts typical goals might be that the forecast errors should
be zero mean (i.e. the forecast is unbiased), or that the forecast errors should have low variance, or both.
For probabilistic forecasts the goal might be that the forecast should maximise the probability
of the observations given the forecast.

In the single-valued forecast case there is often a trade-off between the twin goals
of low bias and low variance, and it may not be possible to reduce one without increasing
the other. The most obvious way to make this trade-off is then to focus on the single
measure of MSE, which is a combination of the mean and the variance.
One case in which such a trade-off occurs in meteorological forecasting
is when we have historical values for a meteorological variable over a number of years,
and each year falls into one of two categories, which we will call A and B.
We assume that next year is going to be an example of category A.
How, then, should we predict the variable?
In the case in which we have many years of historical data for category A, or
where the signal to noise ratio is large, the best single-value forecast is obvious: it consists of the mean
of the historical data in category A.
However, in the case where we only have a small number of historical years in category A,
a larger number in category B,
and the signal to noise ratio is small, how to make the best forecast is
less obvious.
The reason for this is that in this case the means of categories A and B combined may actually
provide a forecast that is nearly as good, or even better (in terms of MSE),
than the mean of category A.
This is because category B
contains more data than category A, leading to a more precise prediction, and this
benefit may be more important than the harm
caused by the data in category B being less representative than that in category A.
Mathematically, the prediction based on category A
has a zero bias but high variance (because of the small amount of data), while the prediction
based on categories A and B together has a non-zero bias, but low variance (because there is more data).
Either prediction could win in terms of MSE, depending on the details (the number of years
of data in each category, and the strength of the signal).
However, a better prediction than either is always a combination of the two predictions.

The way we formulate the hurricane prediction question we have outlined in the introduction
turns it into exactly this kind of problem.
The most naive forecast for hurricane activity for next year is probably the average over
the longest baseline for which we have reliable data.
We will use hurricane count data from 1900, and so in our case that baseline is
1900-2005 (106 years of data). The argument for using such a prediction is that it is based on a
lot of data.
It is clear, however, that the time series of hurricane counts are not stationary during
this period (this issue has been discussed at length in the scientific literature:
see, for example, \citet{goldenberg} or \citet{graysl97}).
In particular, the last 11 years (1995-2005) have
shown more hurricanes than the long-term average,
and the two available physical explanations for this (the Atlantic Multi-Decadal
Oscillation and climate change) both suggest that this increased level will
continue.
This motivates the idea that a
better forecast might be the average over this short baseline. The argument for this
alternative prediction is that it should better capture the phase of whatever cycle or trend
has caused the recent increase in numbers, while the disadvantage is that it is based on less data.

We now put this problem into the categorical forecasting framework outlined above.
Category A is the data for 1995-2005 (11 years), and we believe that 2006 is also
going to be in category A. Category B is the data for 1900-1994 (95 years).
We would expect a prediction based on the average of the hurricane activity in category A to be
unbiased, but to have a high error variance because of the lack of data and the high level of noise.
We would expect a prediction based on the average of the hurricane activity in categories A and B together to be
biased, but to have a lower error variance because more data is being used.
It is unclear a priori which prediction will have the lower MSE.

The goal of this paper is to evaluate these two prediction methods and to
derive the relations that give optimal combinations of them.
To keep the problem as simple as possible, in order to illustrate
the concepts involved, we will make the following straightforward assumptions:

\begin{itemize}

    \item We assume that annual hurricane numbers for 1900-2005 can be modelled as samples
    from poisson distributions

    \item We assume that the mean of these poisson distributions was constant at one level from 1900 to 1994,
    and constant at another level from 1995 to 2005.

\end{itemize}

We could, as is always the case in such a modelling study, make our model more complex.
For instance, it is not correct that mean hurricane rates were constant from 1900 to 1994:
there are clear multidecadal shifts in activity during this period, as mentioned earlier.
And the poisson distribution is not
a perfect fit either, with the data having a slightly larger variance than the mean.
However, we believe that this simple model is a useful addition to the debate about how to
predict future hurricane numbers because it is simple, analytically tractable,
and easy to understand.
It also introduces and allows us to quantify the trade-off between mean error and error variance that lies
at the root of this problem.

We note that we have previously solved the same statistical problem for the case
where the data is normally distributed rather than poisson~\citep{j75}. The context for that study was the prediction of
the impacts of El Ni\~{n}o on US temperatures.

\section{Individual vs. Overall Sample Mean}\label{sectionmean}

In this section we derive the properties of the two most straightforward
ways of making a categorical forecast from two category data: taking the sample mean of both the
categories together, and taking the sample mean within each category.
In the next section we will investigate mixing these two simple predictions
in an optimal way.

Consider random samples from two populations
\begin{eqnarray}
Y_{1,j} \sim \mbox{Pois}(\lambda_i), \ \ j=1,\ldots,n_1 \\\
Y_{2,j} \sim \mbox{Pois}(\lambda_i), \ \ j=1,\ldots,n_2
\end{eqnarray}

Our interest lies in predicting the value ($Y_{1,n_1+1}$)
and the expected value ($E(Y_{1,n_1+1})$)
of a new observation from
(without loss of generality) the first population.
We consider the sample mean of population 1 and the overall sample
mean for populations 1 and 2 as predictors.
First we define the sample mean as:

\begin{equation}
\hat{\lambda}_i = \frac{1}{n_i} \sum_{j=1}^{n_i} Y_{i,j}.
\end{equation}

Our two predictors are then
\begin{eqnarray}
\hat{Y}_{1,n_1+1} &=& \hat{\lambda}_1 \\
Y_{1,n_1+1}^\dagger &=& \frac{n_1}{n_1+n_2} \hat{\lambda}_1 + \frac{n_2}{n_1+n_2} \hat{\lambda}_2
\end{eqnarray}

The properties of these predictors are given below:

\begin{eqnarray}
E(\hat{Y}_{1,n_1+1}-Y_{1,n_1+1}) &=& 0 \quad \mbox{(unbiased)} \\
E(\hat{Y}_{1,n_1+1}-E(Y_{1,n_1+1})) &=& 0 \quad \mbox{(unbiased)} \\
\var(\hat{Y}_{1,n_1+1}-Y_{1,n_1+1}) &=& \left(1+\frac{1}{n_1} \right) \lambda_1\\
\var(\hat{Y}_{1,n_1+1}-E(Y_{1,n_1+1})) &=& \frac{1}{n_1}\lambda_1
\end{eqnarray}

Defining
$\mbox{MSE}_1$ as the mean squared error of predictions of $Y_{1,n_1+1}$,
and
$\mbox{MSE}_2$ as the mean squared error of predictions of $E(Y_{1,n_1+1})$,
we then have:

\begin{eqnarray}
\Rightarrow \mse_1(\hat{Y}_{1,n_1+1}) &=& \left( 1+\frac{1}{n_1} \right) \lambda_1\\
            \mse_2(\hat{Y}_{1,n_1+1}) &=& \frac{1}{n_1}\lambda_1
\end{eqnarray}

\begin{eqnarray}
E(Y_{1,n_1+1}^\dagger-Y_{1,n_1+1}) &=& \frac{n_2}{n_1+n_2} (\lambda_2 - \lambda_1) \quad \mbox{(biased)} \ \\
E(Y_{1,n_1+1}^\dagger-E(Y_{1,n_1+1})) &=& \frac{n_2}{n_1+n_2} (\lambda_2 - \lambda_1) \quad \mbox{(biased)} \ \\
\var(Y_{1,n_1+1}^\dagger-Y_{1,n_1+1}) &=& \left(1+\frac{n_1}{(n_1+n_2)^2} \right) \lambda_1
+ \frac{n_2}{(n_1+n_2)^2} \lambda_2 \\
\var(Y_{1,n_1+1}^\dagger-E(Y_{1,n_1+1})) &=& \frac{n_1}{(n_1+n_2)^2} \lambda_1
+ \frac{n_2}{(n_1+n_2)^2} \lambda_2
\end{eqnarray}

\begin{eqnarray}
\Rightarrow \mse_1(Y_{1,n_1+1}^\dagger) &=& \left(1+\frac{n_1}{(n_1+n_2)^2} \right) \lambda_1
+ \frac{n_2}{(n_1+n_2)^2} \lambda_2 + \left(\frac{n_2}{n_1+n_2} \right)^2 (\lambda_2 - \lambda_1)^2 \\
            \mse_2(Y_{1,n_1+1}^\dagger) &=& \frac{n_1}{(n_1+n_2)^2} \lambda_1
+ \frac{n_2}{(n_1+n_2)^2} \lambda_2 + \left(\frac{n_2}{n_1+n_2} \right)^2 (\lambda_2 - \lambda_1)^2
\end{eqnarray}

Which of these two predictions is better?
The condition under which the overall sample mean is preferable to the
individual sample mean (in terms of $\mbox{MSE}_1$) as a predictor of $Y_{1,n_1+1}$ can now be derived, and is:

\begin{equation}
\mse_1(Y_{1,n_1+1}^\dagger) < \mse_1(\hat{Y}_{1,n_1+1})
\iff (\lambda_2-\lambda_1)^2 < \frac{(n_2 + 2n_1) \lambda_1 - n_1 \lambda_2}{n_1 n_2}
\end{equation}

The same relation holds for $\mse_2$.

\section{General Mixed Predictors}\label{sectionmixed}

We now consider mixing the two forecasts discussed above to create
an optimal categorical prediction.
We consider the general mixed predictor

\begin{equation}
Y_{1,n_1+1}^*(\alpha) = \alpha \hat{\lambda}_1 + (1-\alpha) \hat{\lambda}_2
\end{equation}

Note that
\begin{eqnarray}
\hat{Y}_{1,n_1+1} &=& Y_{1,n_1+1}^*(1), \\
Y_{1,n_1+1}^\dagger &=& Y_{1,n_1+1}^*(n_1/(n_1+n_2))
\end{eqnarray}

The properties of the general mixed predictor are

\begin{eqnarray}
E(Y_{1,n_1+1}^*(\alpha)-Y_{1,n_1+1}) &=& (1-\alpha) (\lambda_2 - \lambda_1)\ \\
E(Y_{1,n_1+1}^*(\alpha)-E(Y_{1,n_1+1})) &=& (1-\alpha) (\lambda_2 - \lambda_1)\ \\
\var(Y_{1,n_1+1}^*(\alpha)-Y_{1,n_1+1}) &=& \left(1+\frac{\alpha^2}{n_1} \right) \lambda_1
+ \frac{(1-\alpha)^2}{n_2} \lambda_2 \\
\var(Y_{1,n_1+1}^*(\alpha)-E(Y_{1,n_1+1})) &=& \frac{\alpha^2}{n_1} \lambda_1
+ \frac{(1-\alpha)^2}{n_2} \lambda_2
\end{eqnarray}

\begin{eqnarray}
\Rightarrow \mse_1(Y_{1,n_1+1}^*(\alpha)) &=& \left(1+\frac{\alpha^2}{n_1} \right) \lambda_1
+ \frac{(1-\alpha)^2}{n_2} \lambda_2 + (1-\alpha)^2(\lambda_2 - \lambda_1)^2\\
            \mse_2(Y_{1,n_1+1}^*(\alpha)) &=& \frac{\alpha^2}{n_1} \lambda_1
+ \frac{(1-\alpha)^2}{n_2} \lambda_2 + (1-\alpha)^2(\lambda_2 - \lambda_1)^2
\end{eqnarray}

In order to find the ``best" mixed predictor we minimize the MSE with respect to $\alpha$

\begin{equation}
\frac{\partial}{\partial \alpha} \left.\mse(Y_{1,n_1+1}^*(\alpha)) \right|_{\alpha = \hat{\alpha}} = 0
\end{equation}

\begin{equation}\label{alpha}
\Rightarrow \hat{\alpha}
= \frac{n_1 n_2 (\lambda_2-\lambda_1)^2 + n_1 \lambda_2}{n_1 n_2 (\lambda_2-\lambda_1)^2
                                        + n_2 \lambda_1 + n_1 \lambda_2}
\end{equation}

Note that this is of the form
\begin{equation}
\hat{\alpha} = \frac{p}{p+q}
\end{equation}

where $p = n_1 n_2 (\lambda_2-\lambda_1)^2 + n_1 \lambda_2$ and $q = n_2 \lambda_1$.

\section{Example}

We now apply the equations derived above to the real case of landfalling hurricanes in the US.
The mean number of landfalling hurricanes from 1900 to 1994 was 1.642105
\footnote{There are various slightly different versions of the data for the number of US landfalling hurricanes
because of (a) different corrections to obvious errors in the data and (b) different definitions of `landfalling'.
Our version is based as closely as possible on the \texttt{SSS} definition from the HURDAT database~\citep{hurdat}.},
while the mean number of landfalling hurricanes from 1995 to 2005 was
2.181818 \footnote{The hurricane season for 2005 is not quite finished yet:
we are assuming that hurricane Alpha will be the last, giving a total of 4 landfalls for 2005.}.

Equation~\ref{alpha} gives a value of $\alpha=0.609$.
The values of bias, error standard deviation and RMSE for predictions made from the 3 forecasts
discussed in section~\ref{sectionmean} and~\ref{sectionmixed} above are given in table 1.
We are principally interested in predicting the \emph{expected} number of hurricanes (rather than just the \emph{number}
of hurricanes), and so we are most interested in SD2 and RMSE2.

\begin{table}[h!]
  \centering
\begin{tabular}{|c|c|c|c|c|c|c|}
 \hline
 model & prediction & mean & SD1 & SD2 & RMSE1 & RMSE2 \\
 \hline
1& 2.182 & 0.000 & 1.543 & 0.445 & 1.543 & 0.445\\
2& 1.698 & 0.484 & 1.482 & 0.127 & 1.559 & 0.500\\
3& 1.971 & 0.211 & 1.503 & 0.276 & 1.517 & 0.347\\
 \hline
\end{tabular}
\caption{Predictions, and properties of those predictions, for three prediction models:
model 1 ($\hat{Y}$) is the `obvious', and unbiased, model based on a short recent baseline,
model 2 ($Y^*$) is the `wrong', and biased, model based on a long baseline
and model 3 ($Y^\dagger$) is the optimal combination of the two.
}
\end{table}

What we see in this example is:
\begin{itemize}

    \item Bias: model 1 (the forecast consisting of the average of the last 11 years)
    has the lowest bias (of zero),
    model 2 (the forecast consisting of the average of the last 106 years) has the
    most bias, and the level of bias from model 3 (the optimum mixture) is in between the two.

    \item Error SD2: model 1 has the \emph{highest} error SD, model 2 the lowest,
    and the level of error SD from model 3 is in between the two.

    \item RMSE2: model 2 has a slightly higher RMSE than model 1, while the RMSE
    for model 3 is the lowest.

\end{itemize}

The behaviour of the bias and error variances are
all exactly as expected.
It is interesting that the short baseline forecast (model 1) beats the long baseline forecast (model 2) in terms
of RMSE, but only just.
The optimal forecast has given us a forecast
with lower RMSE than either of the components, as it should.

\section{Conclusions}

We have considered how to make model-based statistical predictions of landfalling hurricane
numbers a year in advance. The method we use assumes that
we can model annual hurricane numbers as poisson distributions,
with one mean from 1900 to 1994, and another mean from 1995
to 2005. The means are unknown and must be estimated from data. We also assume that 2006 will
come from the same population as the years 1995-2005.

How best to predict the number of hurricanes for 2006
is not obvious because the data is noisy and we have only
a few years of data at the recent higher level of activity.
We derive expressions for the optimal prediction that can
be made using a combination of the data from 1900-1994 with the data from 1995-2005.
Finally we apply these expressions to the real data for these periods to generate our optimal
forecast.

Our forecast is not intended as a genuine prediction of future hurricane activity, since there
are other factors that one would want to take into account, such as the widely held belief
that there have been changes in hurricane activity in the past
(with high activity in the 1940s and 1950s for instance).
This study could be extended to take these complexities into account.
However, our study does illustrate the necessity for taking careful consideration of (a) the need
to define what one is trying to predict and (b) the possibility that biased predictors may out perform
unbiased predictors (in terms of RMSE) when we have little data or the signal to noise ratio is small.

There is one statistical issue that we haven't discussed, and that would merit some further investigation.
The value of $\alpha$ we use in our example is only an \emph{estimate} of the best value of $\alpha$,
and could be rather different from the best value. This is likely to reduce the benefit of making
the optimal combination. One way to understand this better would be to run simulations, as we
did for the corresponding normally distributed problem in~\citet{j77}.

Finally we note that there are other ways that one can approach the same problem.
For example, some might be tempted to use Bayesian statistics,
and others to use bootstrapping methods.
The presentation we have given, however, does seem to be the simplest of the various options.

\section{Legal statement}

SJ was employed by RMS at the time that this article was written.
However, neither the research behind this article nor the writing
of this article were in the course of his employment, (where 'in
the course of their employment' is within the meaning of the
Copyright, Designs and Patents Act 1988, Section 11), nor were
they in the course of his normal duties, or in the course of
duties falling outside his normal duties but specifically assigned
to him (where 'in the course of his normal duties' and 'in the
course of duties falling outside his normal duties' are within the
meanings of the Patents Act 1977, Section 39). Furthermore the
article does not contain any proprietary information or trade
secrets of RMS. As a result, the authors are the owners of all the
intellectual property rights (including, but not limited to,
copyright, moral rights, design rights and rights to inventions)
associated with and arising from this article. The authors reserve
all these rights. No-one may reproduce, store or transmit, in any
form or by any means, any part of this article without the
authors' prior written permission. The moral rights of the authors
have been asserted.

The contents of this article reflect the authors' personal
opinions at the point in time at which this article was submitted
for publication. However, by the very nature of ongoing research,
they do not necessarily reflect the authors' current opinions. In
addition, they do not necessarily reflect the opinions of the
authors' employers.

\bibliography{../bib/jewson}

\end{document}